\documentclass[journal=jctcce,manuscript=article]{achemso}

\usepackage[version=3]{mhchem} % Formula subscripts using \ce{}
\usepackage[T1]{fontenc}       % Use modern font encodings
\usepackage{subcaption}
\usepackage[labelformat=parens,labelsep=quad,skip=3pt]{caption}
\usepackage{graphicx}
\usepackage{gensymb}
\usepackage{float}

\author{Romina Ruberto}
\author{Enrico Smargiassi}
\author{Giorgio Pastore}
\email{pastore@units.it}
\affiliation[University of Trieste]
{Dipartimento di Fisica, Universit\`a di Trieste, Strada Costiera 11, 34151 Grignano (Trieste), Italy}

\title[]
  { The SIRAH force-field and interactions between short DNA duplexes
   }

\begin{document}

%%\begin{tocentry}

%Some journals require a graphical entry for the Table of Contents.
%This should be laid out ``print ready'' so that the sizing of the

%Inside the \texttt{tocentry} environment, the font used is Helvetica
%8\,pt, as required by \emph{Journal of the American Chemical

%The surrounding frame is 9\,cm by 3.5\,cm, which is the maximum
%permitted for  \emph{Journal of the American Chemical Society}
%graphical table of content entries. The box will not resize if the
%content is too big: instead it will overflow the edge of the box.

%This box and the associated title will always be printed on a
%separate page at the end of the document.

%\end{tocentry}

\begin{abstract}
In recent years, short DNA duplexes have been studied as an interesting self-assembling system and a building block for DNA-based nanotechnologies. Numerical simulation studies for the determination of the full phase diagram of short duplexes require, as an input ingredient, a simplified but reliable force-field able to capture the main features of duplex-duplex interaction. We used the coarse-grained SIRAH force field in the implicit solvent approximation to study the interaction between two short duplexes of double-strand DNA as a function of the relative positions and orientations and the salt concentration, in the cases of 8 and 20 base pairs.  We discuss the consequences of our study to determine new simple but qualitatively reliable model potentials.
\end{abstract}

\section{Introduction}

Besides his fundamental role as the molecule of life, DNA is a fascinating material in colloidal science\cite{Pedersen}, due to the possibility
of programming and reproducing with high accuracy and reliability specific sequences of nucleotides. Since the exact alternation of bases 
modifies its 
interaction properties, DNA is currently investigated as a possible building block to control, in a planned way, self-assembly of supermolecular 
structures \cite{hendrikse2019,linko2013enabled,bath2007dna,liedl2007dna}. The phase diagrams of systems made of short fragments of double-strand DNA (duplexes) have been investigated with experimental and theoretical \cite{bellini2011dna,de2011self,nguyen2014self,de2012self}  methods, providing evidence that even short duplexes may display rich and non-trivial phase diagrams. 
In particular, liquid crystal phases (LC)  have been shown to be present even with duplexes as short as  six base pairs\cite{de2011self} (6bp).
Moreover,  a complete understanding of  the interplay between chirality of the duplexes and the presence of chiral mesophases of left and right-handedness 
oligonucleotides remains a challenge\cite{bellini2011dna}.

The possibility of a theoretical understanding of such rich and complex behavior through numerical methods requires realistic
modeling of the interactions
between fragments of double-strand DNA. Fully atomistic calculations in the presence of an explicit description of the aqueous solution medium,
repeated for many concentrations and thermodynamic states, are still out of reach for the present computational capabilities.
Even coarse-grained (CG) models are not suitable for extensive samplings of the thermodynamic phase space. Still, they may play an essential role in transferring the all-atoms description into the development of more simplified but physically correct and computationally efficient force fields.

Although CG models for interactions between biological molecules have been developed over many years, specific CG interactions for DNA have appeared 
only recently \cite{savelyev2010chemically,uusitalo2015martini,ouldridge2011structural,doye2013coarse,SIRAH1,machado2011hybrid,dans2013assessing}(see a recent review for a comprehensive list\cite{noid2013perspective}). 
Most of the existing tests and applications have been restricted to intramolecular interactions, which is necessary to study phenomena like pairing between complementary single strands or intra-molecular defects.  Much less is known about the capabilities of such force-fields to  describe the inter-duplex interactions. Still, they are key ingredients for modeling recently proposed DNA-designed colloidal systems. On the one hand, LC phases strongly depend on the effective interactions between duplexes; on the other hand, the analysis of recent experiments \cite{zanchetta2010right,frezza2011right} has shown the need for more accurate force-fields, beyond the simplest  models, like hard rods decorated with attractive sites\cite{de2012self}.

A first successful and accurate CG model for DNA was proposed in recent years by Pantano and coworkers\cite{SIRAH1}, which named it  SIRAH (South-American Initiative for a Rapid and Accurate Hamiltonian) force field.
More recently, other CG models have been presented  \cite{uusitalo2015martini,ouldridge2011structural,doye2013coarse}. In some cases, they differ for the level of coarse-graining; in other cases, the choice between comparable levels of coarse graining  may be driven by differences in the accuracy of reproducing some physical 
properties or by practical considerations related to the level of integration in various molecular dynamics packages or with other force fields. 

In the present paper, we have studied the energy landscape and the resulting forces between two short duplexes of DNA described by SIRAH force field. 
We have calculated energy and force on the two double-strands of size 8bp and 20bp for a representative set of configurations in the implicit solvent approach (i.e., in an approach such that the effect of the ionic solution is fully absorbed into modifications of the bare DNA-DNA interactions). For a small subset of configurations, we have also performed a few calculations with the explicit solvent model where water and salt ions are treated at the same coarse-grained level of the DNA. These calculations were intended to provide an assessment of the role of the solvent. The present results will help to develop new rigid duplex-duplex interaction models, suitable for extensive numerical simulations of phase diagrams.

The paper is organized as follows.  In Section 2, we briefly describe the force field model and how we have used it in our calculations.
In Section 3, we present our detailed analysis of the energy landscape and forces.  In the conclusions, we have collected a short summary and a few comments on our results.

\section{The model}

Numerical simulation of DNA is a complex challenge. Space scales go from interatomic distances within its constituent molecules to macroscopic lengths 
(of the order of one meter) for the unswelled double helix of human DNA\cite{SIRAH1}. Similarly, time-scales of all the possible 
dynamical processes span 
more than twenty decades. Thus, more than a unique modeling strategy, it has been recognized useful to develop several methods, 
each adapted to a particular 
level of resolution, going from quantum chemistry accuracy (molecular level) to elastic mesoscopic models, passing trough mixed 
Quantum Mechanics/Molecular Mechanics,  atomistic and coarse-grained levels of description.

Particle-based coarse-grained (CG) models are based on  replacing groups of multiple atoms by effective beads interacting in such a way to maintain the geometry of the original groups (see ref. \cite{Voth} for a comprehensive review). Models differ in the number of atoms represented by a single bead. If combined with a similar CG modeling of the solvent, particle CG models can be quite useful in reducing the number of independent degrees of freedom, while keeping good accuracy in the description of structure and energy.
DNA is clearly a molecule suitable for CG modeling and different research groups have developed CG models for numerical simulations \cite{Potoyan_review}. Among different CG models, the SIRAH  force field\cite{SIRAH1} looks an interesting model, flexible enough to be used in combination with explicit or implicit solvent description and accurate enough to be considered a possible benchmark for more approximate model hamiltonian.

\subsection{The SIRAH force field}

The SIRAH force field for DNA uses six beads for each nucleic base. Each bead is at the same position as an atom in the atomistic reference structure, thus allowing to reconstruct quite easily the atomic positions if required. A partial electrostatic charge is added to the beads so that each nucleotide carries a unitary negative charge to ensure Watson-Crick electrostatic recognition. Moreover, the resulting electric  dipole distribution is compatible with all-atoms models \cite{SIRAH1}. The model also retains the identity of minor and major grooves, as well as the 
5$^{\prime}$ to 3$^{\prime}$ directionality of DNA helices.
In addition to the electrostatic term, bead-bead interactions include\cite{SIRAH1}: i) a harmonic bond-stretching term, ii) a harmonic angle-stretching term, iii) a dihedral torsional barrier, and iv) an effective $12-6$ Lennard-Jones interaction, which is mainly responsible, through its repulsive part, for bead-bead (and in turn nucleotide-nucleotide and finally DNA-DNA) excluded volume effects. Parameters for bead-bead interactions are listed in the original paper\cite{SIRAH1}.

Within the  SIRAH model, the solvent can be treated at two different levels.
The implicit solvent level \cite{SIRAH1} takes care of hydration and ionic presence effects through the Generalized Born model\cite{hawkins1996parametrized}, while the explicit solvent model restores the presence of solvent and ionic degrees of freedom through a coarse grained description of water molecules (WAT FOUR)\cite{darre2010another} and hydrated ions. In WAT FOUR, groups of 11 water molecules are represented by a group of four  tetrahedrally interconnected beads, hence the name WAT FOUR (WT4). Since each bead carries an explicit partial charge, WT4 liquid generates its own dielectric permittivity without the need to impose a uniform dielectric medium. The model reproduces several common properties of liquid water and simple electrolyte solutions\cite{darre2010another}. Similarly, each group of cation (Na$^+$,K$^+$,Ca$^{++}$) plus its hydration sphere is modeled by a CG molecule. A complete list of all parameters can be found in the original paper\cite{darre2010another}. The CG model for the solvent was extensively tested by its authors. Taking into account the intrinsic limitations of a CG model where each bead represents 11 water molecules, it  provides a satisfactory overall description of  the properties of water as compared with  results obtained by using well established atomic models (TIP4P, SPC)\cite{darre2010another}.

The solvated ions are represented by CG particles corresponding to a central ion surrounded by six water molecules always attached to them (i.e., roughly considering an implicit first solvation shell). Therefore, their masses are the sum of the ionic mass plus that of six water molecules. Partial charges are set to unitary values. The van der Waals radii are adapted to match the first minima of the radial distribution function (RDF, also known as g(r)) of hydrated ions as obtained from neutron diffraction experiments. The well depths have the same values as in the WT4 beads. This  ensures that, when a WT4 molecule touches a CG ion, it interacts implicitly with its first solvation shell. 

\subsection{Calculations with the SIRAH model}

\begin{figure}[H]
\centering
\vspace*{50mm}
\includegraphics[scale=.65]{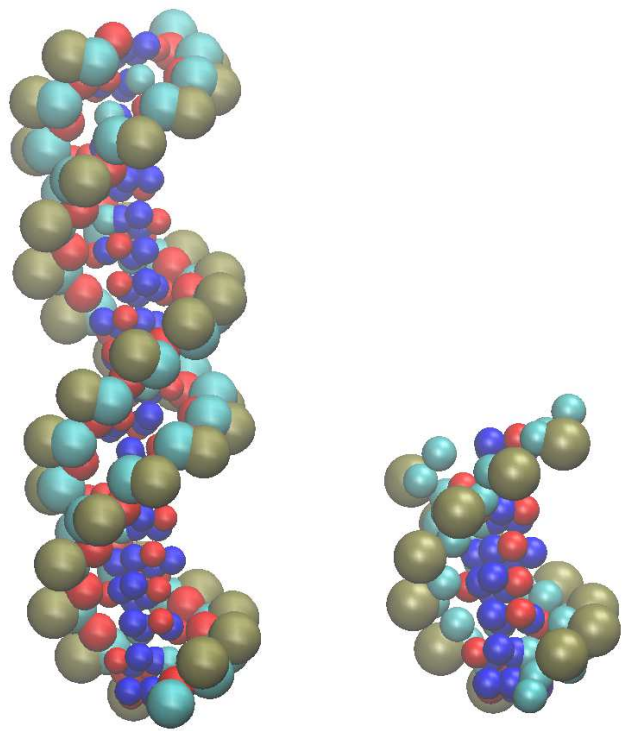}%{./figure_cap/vmdscene_white.eps}
  \caption{The two duplexes used in our calculations. On the left 20pb-dsDNA and on the right the
  8pb-dsDNA fragments.
}
  \label{fgr:duplexes}
\end{figure}

All our calculations have been performed for  an isolated pair of each of the two short duplexes (20 base pairs (bp) and 8 bp-long double stranded (ds) 
DNA). Their SIRAH coarse-grained representation is illustrated in Fig. \ref{fgr:duplexes}.

%1   verra' numerata come fig 1

The 20-mer duplex is based on the sequence 
5$^{\prime}$ -(C$_1$A$_2$T$_3$G$_4$C$_5$A$_6$T$_7$G$_8$C$_9$A$_{10}$T$_{11}$G$_{12}$C$_{13}$A$_{14}$\-T$_{15}$G$_{16}$C$_{17}$A$_{18}$T$_{19}$G$_{20}$
)-3$^{\prime}$ , indicated as S$_{AA2}$ by Machado et al.\cite{machado2011hybrid}. The shorter 8 bp duplex corresponds to the first 8 nucleotides of the 20-mer. Its choice was mainly motivated by the possibility of having the same terminal bases for both sizes. In all cases the model building procedure starts from the Cartesian coordinates of structures containing all the atoms, built in the canonical B-form of DNA33 using the NAB utility of AMBER\cite{pearlman1995amber,case2005amber}, and then the SIRAH scheme is used for mapping from atomic level to CG level nucleobases by simpy removing and renaming the corresponding atoms. All the bonding parameters used in the model are contained in Tables 1 and  2 of the original description of the force field\cite{SIRAH1}.

As  anticipated in the introduction, the main aim of the present study  has been to obtain information about the SIRAH energy landscape of duplex-duplex interaction in the case of two isolated duplexes at fixed structure of the duplex. 
Such constraint is quite artificial, since the real interaction between helices certainly implies some atomic relaxations. 
However, we notice that our duplexes are very short and the main effect we are interested is the deviation from a cylindric symmetry approximation due to the three-dimensional structure of the double helix. We believe that for this aim a rigid model of the double helix is sufficient.

To understand the extent the solvent is perturbed by the presence of the duplexes, we also performed some calculations with the explicit solvent model. In this case, 
water CG molecules and hydrated CG ions were chosen to simulate a 1M aqueous solution of NaCl. The simulation cell was a truncated octahedron to optimize the number of solvent particles\cite{allen2017computer}. The volume was $426.1~nm^3$.
In the case of explicit solvent calculations, the simulation box for the two 20-mer case contained $6036$ coarse-grained particles. In addition to the two DNA duplexes, there were $1431$ coarse-grained water molecules groups, $76$ hydrated sodium ions and $48$ hydrated chloride ions.
Coarse-grained water molecules were described by   the so-called WAT FOUR  model\cite{darre2010another}.
Molecular dynamics calculations were performed with Gromacs\cite{Gromacs} package. 

In the case of implicit solvent calculations, an extensive sampling of configurations of the system has been performed, varying distance, mutual orientations and salt concentration.  Hydration and ionic strength effects are implicitly taken into account, through their effect on electrostatic interactions  using the generalized Born (GB) model\cite{hawkins1996parametrized} for implicit solvation as implemented in AMBER\cite{pearlman1995amber,case2005amber}.  
The Born effective radii were fixed to $0.15~nm$ for all superatoms. The maximum distance between atom pairs considered in the pair-wise summation involved in calculating the effective Born radii is set to $1~nm$.  
Non-bonded interactions are calculated up to a cutoff of $8.6~nm$ within the GB approximation and the salt concentration varies from 0.15 M to 2.0 M. 

\begin{figure}
%  As well as the standard float types \texttt{table}\\
%  and \texttt{figure}, the class also recognises

%  \texttt{scheme}, \texttt{chart} and \texttt{graph}.
\centering
\includegraphics[scale=0.4]{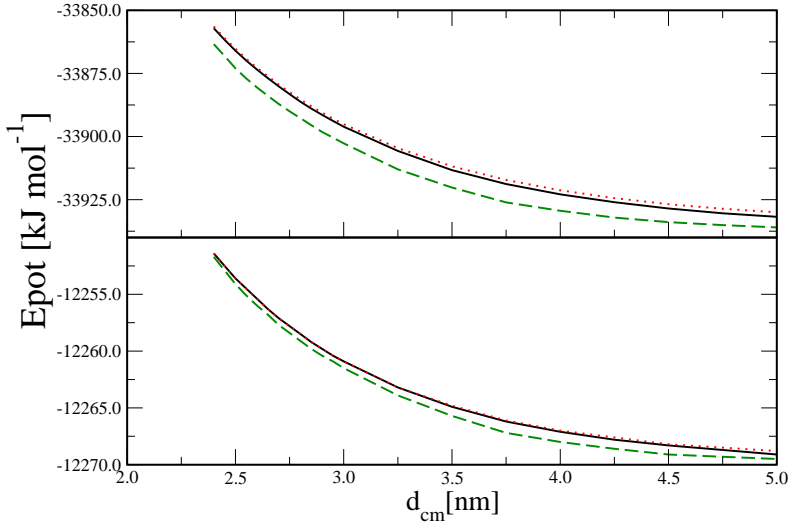}
  \caption{Nonbondend interactions cutoff-effects for two parallel dsDNA 
in the implicit solvent model, at salt concentration 0.15M. Different curves correspond to the following values of the interaction cutoff distance:
green long-dashed: 3.6nm,
black solid-line: 4.8nm,
red dotted line:  6.4 nm. Calculations with a cutoff distance of 8.6 nm are indistiguishable on the scale of the figure from the red dotted points.
Top panel: 20bp-dsDNA, 
Bottom panel: 8pb-dsDNA. d$_{cm}$ indicates the distance between the centers of mass of the two duplexes.}
%nel dettaglio uno dei due e' ruotato di 20 gradi attorno all'asse z
\label{fgr:cutoff}
\end{figure}

Tests on the dependence of results on the choice of the cutoff are shown in Fig. \ref{fgr:cutoff}.
Although quantitative differences are still visible between calculations with a cutoff at $3.6~nm$ and  $4.8~nm $, this last value can be considered already close to numerical convergence and was used in all the calculations. We also checked that this cutoff ensures a smooth behavior of forces for all the configurations of the present study.
\bigskip
\bigskip
\bigskip
%2

\section{Results}
\subsection{Duplex-duplex interaction}

Calculations with the implicit solvent model are straightforward. Among the many different configurations we have studied, here we discuss a representative set to illustrate the main features of the inter-duplex interactions.

\begin{figure}
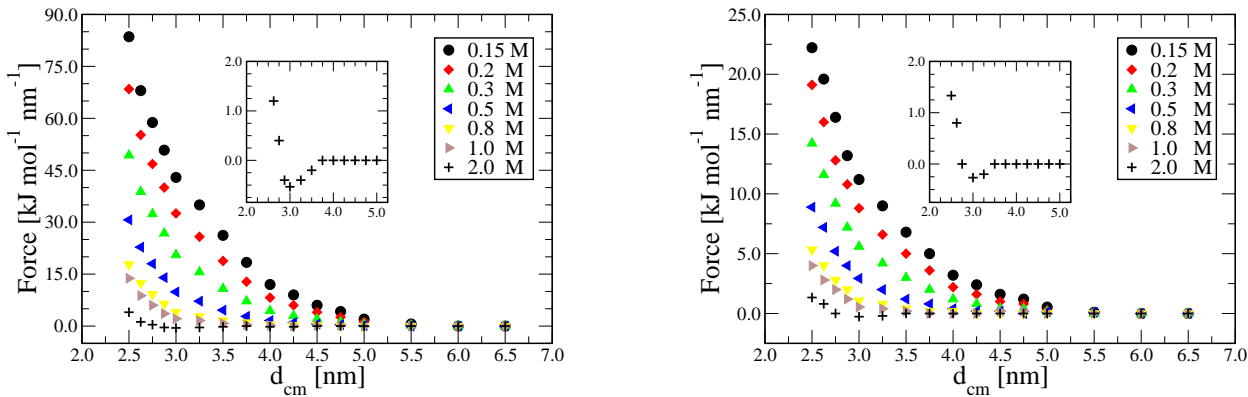

\vspace*{10mm}
\begin{subfigure}[h]{0.45\linewidth}
\includegraphics[width=\linewidth]{./figure_cap/aaFig_3.eps}
%\caption{Image A}
\end{subfigure}
\hfill
\begin{subfigure}[h]{0.45\linewidth}
\includegraphics[width=\linewidth]{./figure_cap/Fig_3newnew.eps}
%\caption{Image B}
\end{subfigure}%
\caption{Intermolecular force curves between two parallel dsDNA 20pb helices (on the left) and two parallel dsDNA 8pb helices (on the right) in the
implicit solvent simulations at different salt concentrations, as a function of center of mass distances (d$_cm$). In the insets: points for 2M concentration on an enlarged scale.}
  \label{fgr:par20and8conc}
\end{figure}

In Fig. \ref{fgr:par20and8conc} we see the variation of the inter-duplex total force between two 20bp (left panel) and two 8bp (right panel)  dsDNA helices at salt concentration ranging 
from  0.15M to 2.0M. 

%3

%\begin{figure}[H]
%  As well as the standard float types \texttt{table}\\
%  and \texttt{figure}, the class also recognises\\
%  \texttt{scheme}, \texttt{chart} and \texttt{graph}.
%\centering
%\vspace*{10mm}
%\includegraphics[scale=0.35]{./figure_cap/aaFig_3.eps}
%  \caption{Intermolecular force curves between two parallel dsDNA 20pb helices in the
%implicit solvent simulations at different salt concentrations, as a function of center-center distances. In the inset: points for 2M concentration on an enlarged scale.}
%  \label{fgr:par20conc}
%\end{figure}
As expected, by increasing the salt concentration, screening of the Coulomb interaction becomes increasingly effective, reducing the strength and range of the repulsion significantly. In all the cases, repulsions are quite short-range, and even for the 0.15M case, they become negligible beyond $5.5$ nm. More important, at 2M concentration, a small attraction appears  with both the duplexes. In the insets of Fig.  \ref{fgr:par20and8conc}, the effect is shown on an enlarged scale.

In the presence of multivalent-ion in solution, the possibility of a range of attractive interactions is well established in experiments and computer simulation\cite{Mu,Kandu,Andresen}.
Less is known for  the case of mono-valent ions, although recent numerical simulations\cite{Luan} predict the effect even for NaCl at concentrations higher than 1M. 
Interestingly, the implicit solvent method reproduces  such  behavior. Experiments on the interaction at high NaCl concentration could provide a direct check for such a prediction.

\begin{figure}
%  As well as the standard float types \texttt{table}\\
%  and \texttt{figure}, the class also recognises\\
%  \texttt{scheme}, \texttt{chart} and \texttt{graph}.
\vspace*{10mm}
\centering
\includegraphics[scale=0.4]{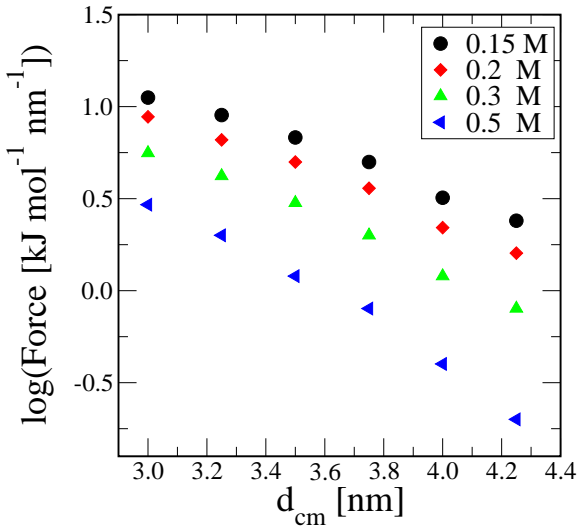}
  \caption{
Plot of the logarithm of the intermolecular force  between two parallel dsDNA 20pb helices, at four different salt concentrations,  as a function of the double strand inter-axial distance, 
in the range of close intermolecular distances. 
}
\label{fgr:logscale}
\end{figure}

An analysis of the force in the logarithmic scale (see Fig.\ref{fgr:logscale}), allows comparison with the experimental data analyzed by Podgornik et al.\cite{podgornik94} in the same range of distances. Their main conclusion was that the interactions at distances below $3$ nm are influenced by solvent layering effects, while for larger distances, at least up to $3.5$ nm,  the repulsion is almost exponential with a decaying length depending on salt concentration. In the implicit solvent model, there is no direct modeling of the solvent; therefore we do find a pretty good exponential decaying of the interactions starting from about $2.5$ nm. Interestingly, the implicit solvent model parameters take care of the salt concentration effect, resulting in a concentration dependent repulsion with screening length which reasonably compares  with the values obtained by Podgornik et al.\cite{podgornik94}  from the direct analysis of experimental data.
Indeed, between  $0.15$ M and $0.5$M we find decaying lengths ranging from $1.63~nm$ to $1.23~nm$, while Podgornik et al.\cite{podgornik94} in their table 1, provide values going from $1.32~nm$ at $0.2$M to $0.76~nm$ at $0.6$ M. %4
%\begin{figure}
%  As well as the standard float types \texttt{table}\\
%  and \texttt{figure}, the class also recognises\\
%  \texttt{scheme}, \texttt{chart} and \texttt{graph}.
%\centering
%\vspace*{10mm}
%\includegraphics[scale=0.4]{./figure_cap/Fig_3newnew.eps}
%  \caption{ 
%Intermolecular force curves between two parallel dsDNA 8pb helices in the
%implicit solvent simulations at different salt concentrations. In the inset: points for 2M concentration on an enlarged scale.
%}
%  \label{fgr:par8conc}
%\end{figure}

%4

We have evaluated the anisotropy of the side-by-side interaction between two parallel duplexes has been evaluated. In agreement with ``primitive model'' calculations \cite{AllahyarovLowen} and 
the theoretical predictions by Kornyshev and Leikin\cite{KornyshevLeikin,KornyshevLeikin2},  we find a rapidly decaying dependence of the interaction energy on the mutual azimuthal angle. 
According to Kornyshev and Leikin\cite{KornyshevLeikin,KornyshevLeikin2}, the dominant dependence of the interaction energy on the azimuthal angle should be of the form 
\begin{equation}
E = E_0(1 + \alpha_1 cos(\phi -\phi_0)).
\end{equation}
Already at a inter-duplex distance of 2.4 nm such a dependence is able to provide a very good description of the angular dependence with an azimuthal modulation amplitude $\alpha$ as small as $0.04$, reducing to $0.01$ at a distance of $4.0$ nm and becoming negligible at $7.0$ nm.
%5
\begin{figure}
%  As well as the standard float types \texttt{table}\\
%  and \texttt{figure}, the class also recognises\\
%  \texttt{scheme}, \texttt{chart} and \texttt{graph}.
\centering
\vspace*{15mm}
\includegraphics[scale=0.4]{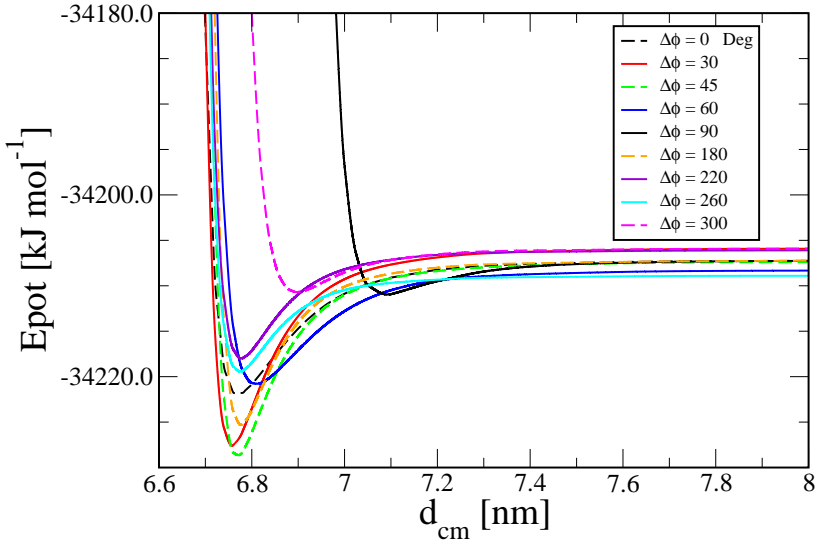}
  \caption{ 
Intermolecolar potential energy curves between two head-tail aligned
dsDNA 20pb molecules, in the implicit solvent simulations at salt
concentration 0.8M, at different rotation angles  of one of them around its double helix axis.
}
\label{fgr:HT}
\end{figure}

The head-tail interaction plays an important influence on the possible phase diagrams of short duplexes systems\cite{bellini2011dna}. Moreover, the possibility and the conditions for end-to-end  DNA association have been less studied in the literature than side-by-side interactions.
Depending on the presence or not of a head-tail 
attraction, it would be possible to favor liquid crystal phases or even reversible chaining between duplexes. Fig. \ref{fgr:HT} shows that there is an 
important angular dependence of such an attraction, which is strongly reduced at some angles. 
The main message from these calculations is again that 
the real three-dimensional interaction between duplexes could be approximated with cylindrically symmetric models only at 
a crude level. 

%6
\begin{figure}
%  As well as the standard float types \texttt{table}\\
%  and \texttt{figure}, the class also recognises\\
%  \texttt{scheme}, \texttt{chart} and \texttt{graph}.
\centering
\includegraphics[scale=0.4]{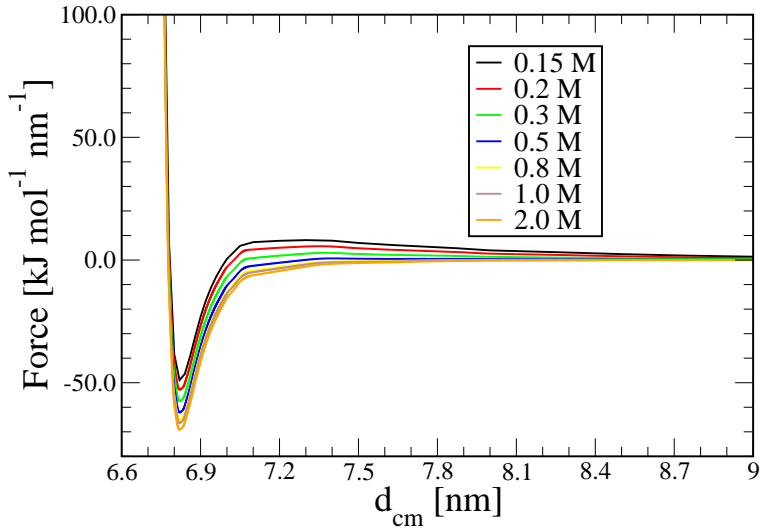}
  \caption{
Intermolecular force curves between two head-tail aligned dsDNA 20pb
molecules, one of which with an azimuthal angle equal to 260\degree, in the
implicit solvent simulations at different salt concentrations.
}
  \label{fgr:HTconc}
\end{figure}

Figure \ref{fgr:HTconc} is an example of the variation of the head-tail force between two aligned 20pb duplexes, as a function of concentration.
The effect of adding salt in the range from $0.15M$ to $2M$ is a regular increase of the attraction of about $40$\%.

%7
\begin{figure}
%  As well as the standard float types \texttt{table}\\
%  and \texttt{figure}, the class also recognises\\
%  \texttt{scheme}, \texttt{chart} and \texttt{graph}.
\centering
\includegraphics[scale=0.4]{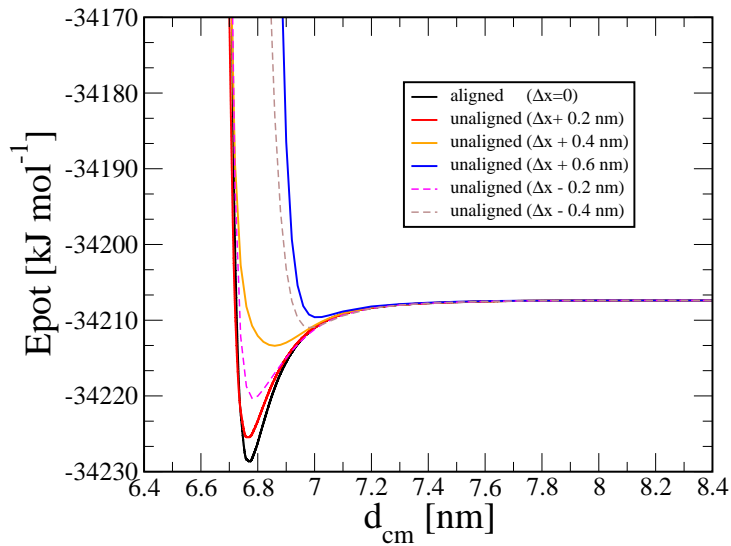}
  \caption{
Potential energy of two 20bp duplexes in the head-tail configuration. The parallel molecular axis has been shifted by $\Delta x$ along the x-axis. 
}
  \label{fgr:disal}
\end{figure}

Finally, 
in Fig. \ref{fgr:disal}, we show the dependence of the interaction energy  on the distance  molecular axes, in a paralel head-tail configuration of two 20pb duplexes. Displacement $\Delta x = 0$ means that the two duplexes are aligned.. As it can be seen by the curves, even a small displacement from a common axis results in a strong reduction of the attractive force.

\subsection{Effect of the solvent}
In the case of explicit solvent calculations, we have explored the possibility of obtaining a meaningful average description of the solvent around the duplexes. We have averaged the system over times of the order of 400 ns, arriving to 1 $\mu$s, in a few cases. 
Although simulation times were long enough to reach a good equilibration of the solvent, it turned out very difficult to obtain a 
reasonable equilibration of the counterions. As it is well known, sodium cations tend to stick  in the major grooves of the 
double helix and the coarse grained hydrated sodium ions of the SIRAH model nicely reproduce such behavior. 
However, the strength of attraction results in very long residence times. During our simulations, the average residence time of the 
hydrated sodium atoms adsorbed on the duplex can be estimated longer than  300 ns, which means that even the longest runs corresponding to  1 $\mu$s provide 
quite a poor statistics. 
Even averaging over randomly chosen initial configuration, we could not get reliable results, due to the large difference of probability of the relevant configurations. 
We conclude that to obtain accurate  quantitative information about the interactions between duplexes in the presence of counterions, it would be necessary  
to increase more than two orders of magnitude the number of time  steps, or, as an alternative, special methods to 
treat rare events should be adapted to the present problem. 

Nevertheless, even if the difficulty of a good statistical sampling 
prevented to obtain quantitative data on the interactions to be compared with the implicit solvent calculations, we notice that the quite noisy averages we obtained are compatible with the results for the implicit solvent model. Furthermore, reliable results for the average solvent density around the duplexes could be obtained.

\begin{figure}
%  As well as the standard float types \texttt{table}\\
%  and \texttt{figure}, the class also recognises\\
%  \texttt{scheme}, \texttt{chart} and \texttt{graph}.
%\centering
   \begin{subfigure}{5cm}
    \centering
    \resizebox*{5cm}{!}{\ \includegraphics{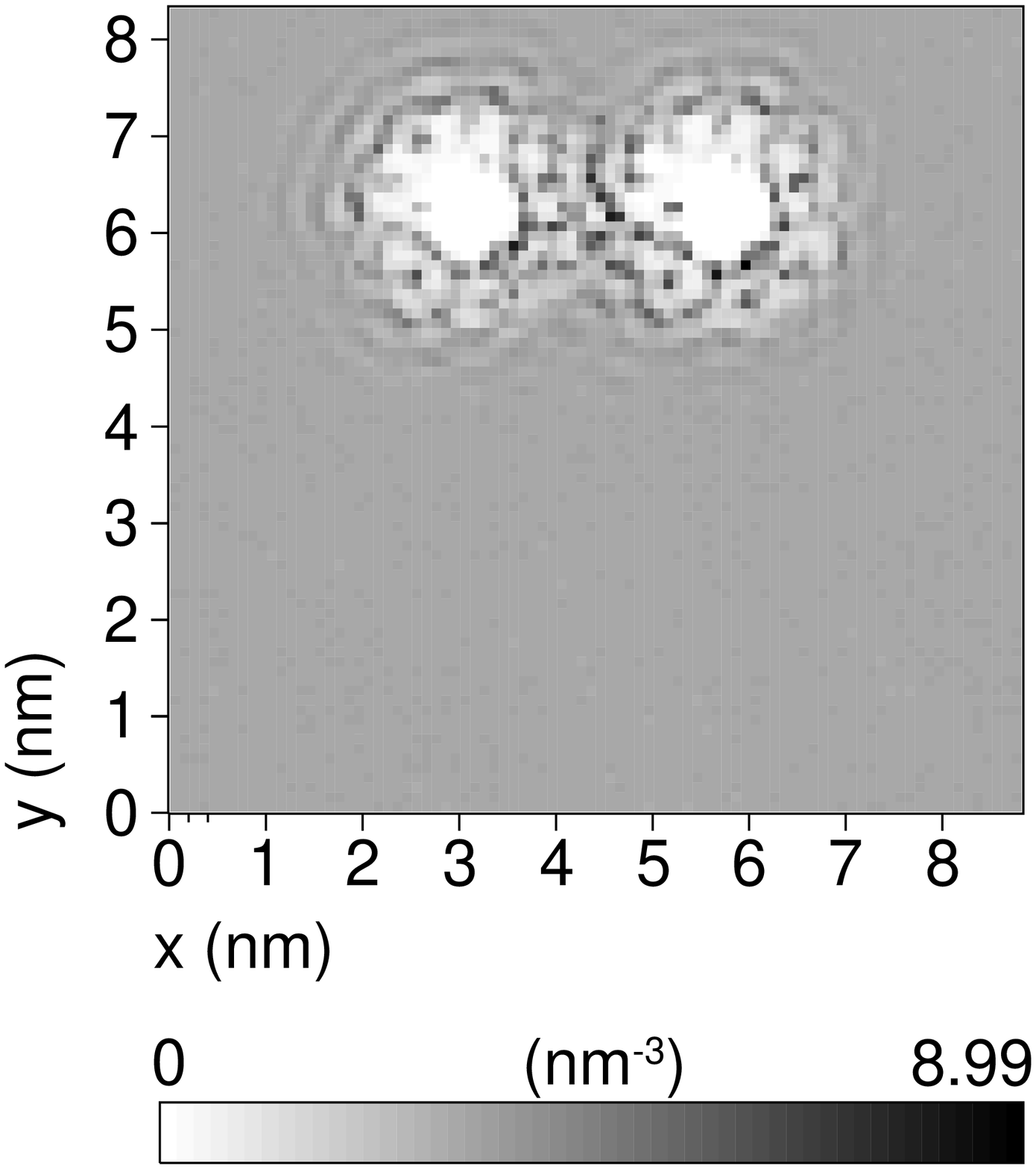} } \hspace{5pt}
    \end{subfigure}
   \begin{subfigure}{5cm}
    \centering
    \resizebox*{5cm}{!}{\ \includegraphics{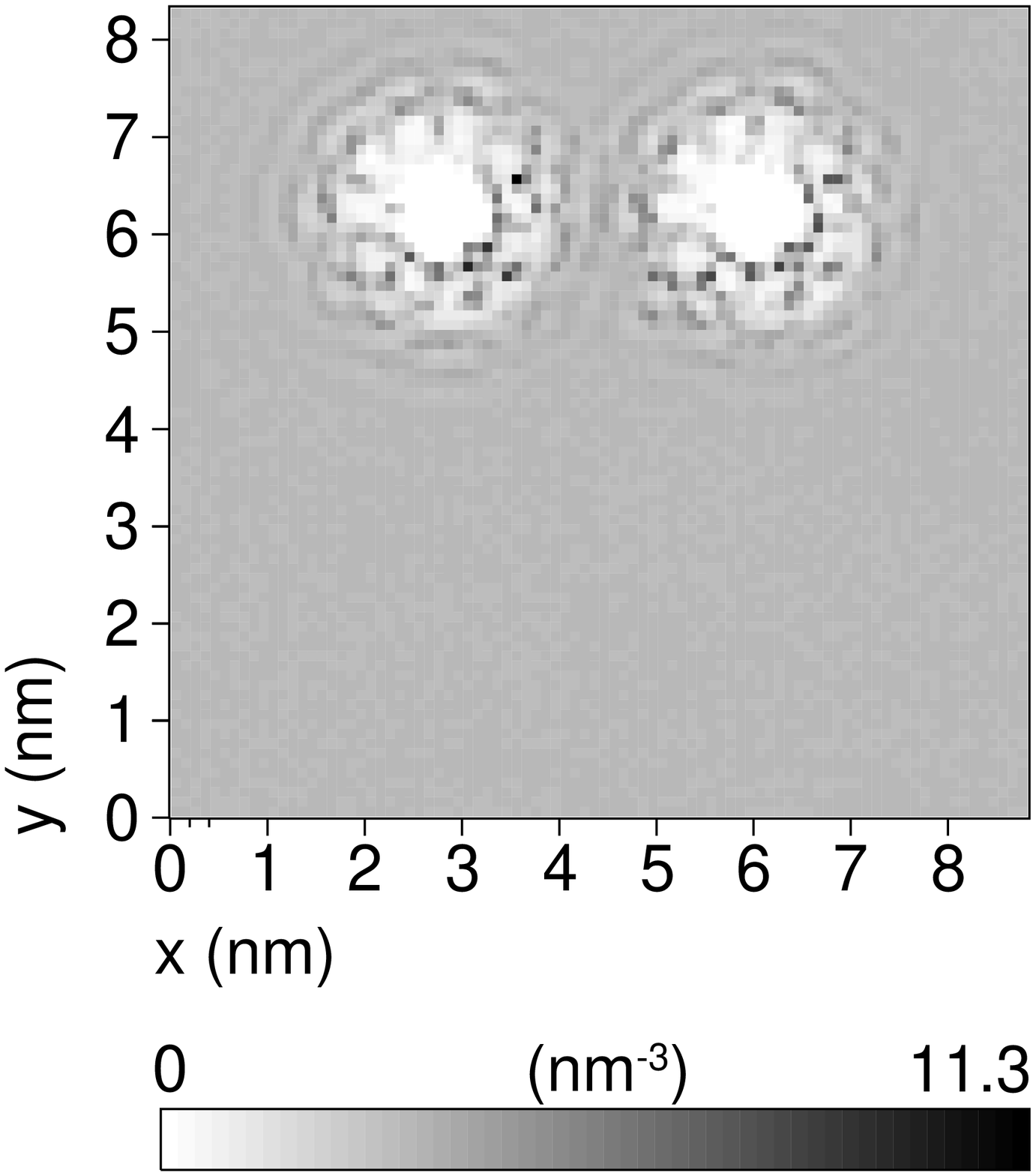} } \hspace{5pt}
    \end{subfigure}
   \begin{subfigure}{5cm}
    \centering
    \resizebox*{5cm}{!}{\ \includegraphics{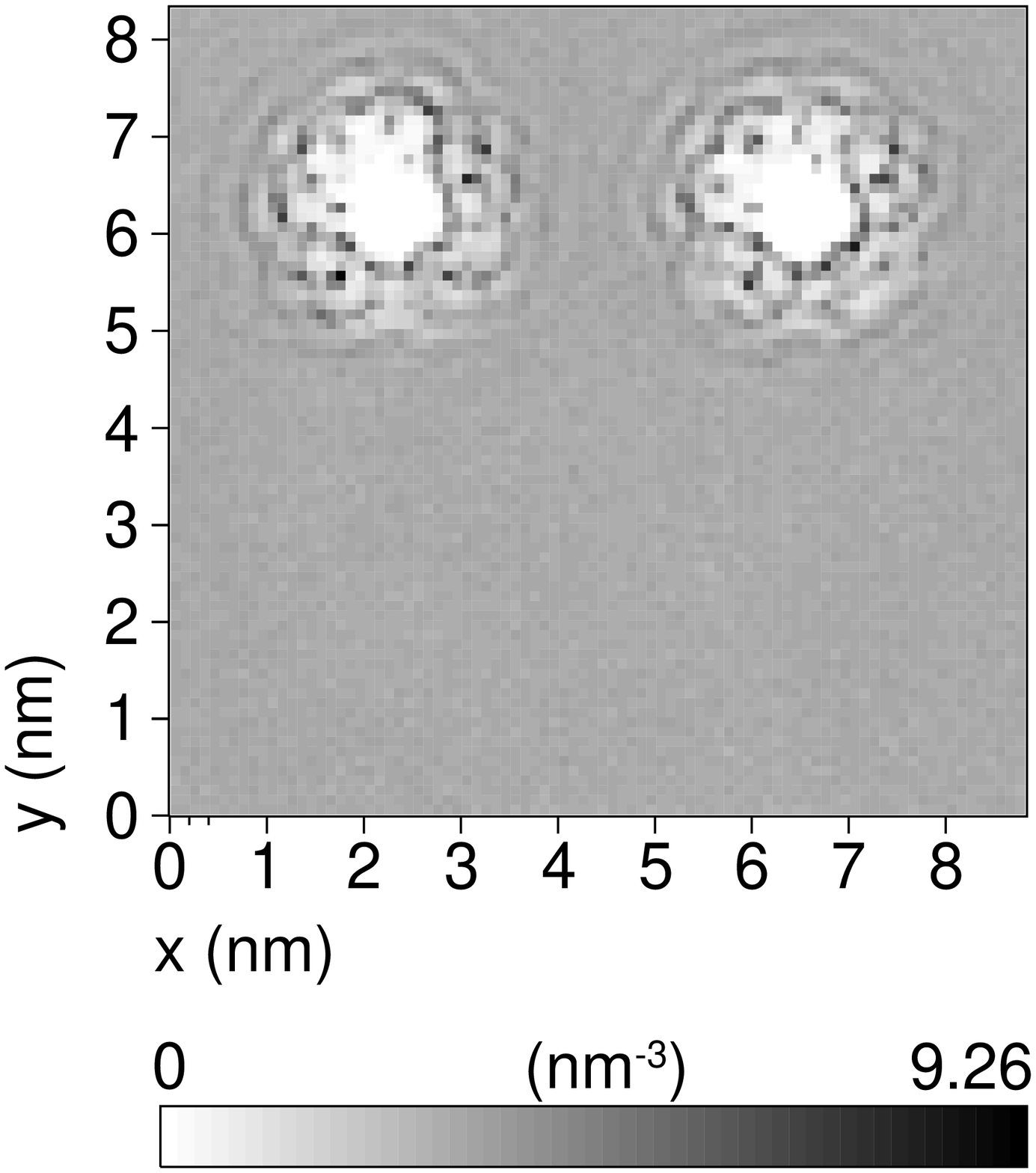} } \hspace{5pt}
    \end{subfigure}
     \caption{
       2D planar WT4 density maps around two parallel dsDNA 8pb molecules at
        interaxial spacing: 2.625 nm (left panel), 3.25nm (central panel), 4.25nm (right panel),
        in the explicit solvent simulations at salt concentration 0.15M.
      }

  \label{fgr:density}
\end{figure}

We present in 
%
%%8
%\begin{figure}
%  As well as the standard float types \texttt{table}\\
%  and \texttt{figure}, the class also recognises\\
%  \texttt{scheme}, \texttt{chart} and \texttt{graph}.
%\vspace*{15mm}
%\centering
%\includegraphics[scale=0.4]{./figure_cap/aaFig_9.eps}
%  \caption{
%Comparison between intermolecular force, as a function of the distance  between two parallel
%dsDNA 8pb molecules, obtained from implicit (solid-line) and explicit solvent 
%(diamonds with the statistical error-bar) simulations at salt concentration
%0.15M.
%}
%\label{fgr:comp_en}
%\end{figure}
%
%In Fig. \ref{fgr:comp_en} we see the comparison between the intermolecular forces between two short duplexes (8bp) in the parallel configuration as obtained from explicit solvent calculations (points with statistical error bars) and implicit solvent (continuous curve). The agreement is only semiquantitative. We look at the scatter of the explicit solvent points, larger than the statistical error bar, as a fingerprint of the systematic error connected to the non-ergodicity due to a very slow counterion dynamics. Even within the large uncertainty of data, the implicit solvent curve is compatible with explicit solvent results.
%Moreover, even if not suitable for quantitative comparison, explicit solvent simulations do provide some useful information. In particular, it is possible to analyze the water density around the two duplexes.
%A typical result of such analysis is shown in 
Fig. \ref{fgr:density} the average density of CG water molecules in a slab of  2 nm perpendicular to the axes of two parallel 8bp double strands. The density values are represented by using a grey-scale code (darker grey means higher density) for three distances between duplexes.
It is clearly visible a modulation of the CG water profile extending around each duplex on a radial length scale of about 1 nm.
Although the previously mentioned ergodicity problems make it difficult to get a direct, reliable quantitative estimation of the effect on duplex-duplex interaction, it is reasonable to expect that such solvent modulation could have qualitative effects on the 
DNA-DNA interaction which are not presently included in the implicit interaction models.

%9

\section{Conclusions}

We have presented a nuerical study of the SIRAH CG model energy landscape for the interaction of two short duplex of DNA with implicit solvation. The main aim of the present study was to use a quite refined CG model to provide qualitative and quantitative input for more approximate interaction models required to study condensed phases of DNA double strands and simulate properties of DNA-based colloids. A certain number of potentials of this kind have been introduced and used in recent years. We hope that our study, based on an accurate CG model,  could help to build intuition about some features 
missing in more crude approximations.

Explicit solvent calculations provided information about the size of the hydration region. Due to the CG model for water, our results could somewhat overestimate the effect, but are in agreement with the theoretical understanding of the DNA-DNA interaction\cite{podgornik94}. The presence of modulations in the solvent density between two duplexes is expected to have some effect on the short range interaction. 

The implicit solvent calculations of the present study have provided thorough information about the characteristics of duplex-duplex interactions.
In particular, we have found an important dependence on salt concentration, with indication of a possible lateral attraction above $2M$.
Dependence on azimuthal rotation is coherent with theoretical predictions, and its repulsive nature implies that averaging interactions over the angle should result in large effective size of double strands.
Head-tail interactions, critical ingredients for the possibility of LC phases, are quite sharply concentrated along the molecular axis and strongly depend on the relative azimuthal angle. This last fact implies that once a chain of duplexes is formed, relative rotations of individual duplexes are strongly suppressed. On the other hand, such angular dependence may act as an additional kinetic constraint for chain formation.
Salt concentration plays an important role, with a possible development of a weak lateral attraction at the highest concentrations.

The picture obtained from our analysis aims to be a solid starting point for future progress in developing simple but realistic model interactions to study condensed phases and phase diagrams of small duplexes and DNA-coated colloids. 

\begin{acknowledgement}

The authors thank Rudi Podgornik for useful comments and suggestions.
GP acknowledges MIUR for financial support through the grant  2010LKE4CC\_004.
RR thanks S. Pantano for useful discussions.

\end{acknowledgement}

\bibliography{DNApaperC}

\end{document}